\documentclass{sig-alternate}

\usepackage{url}
\usepackage{booktabs}
\usepackage{hhline}
\usepackage{subcaption}
\usepackage{color}
\usepackage{array}
\usepackage{amsmath}
\usepackage[utf8]{inputenc}
\usepackage{tikz}
\usepackage{ upgreek }
\usepackage{multirow}
\usepackage{enumitem}
\usepackage{multirow}
\usepackage{algorithm}
\usepackage{algorithmicx}
\usepackage[noend]{algpseudocode}
\usepackage[normalem]{ulem}
\usepackage{setspace}

\DeclareMathOperator*{\argmin}{arg\,min}

\definecolor{alertcolor}{rgb}{0.2, 0.7, 0.4}
\definecolor{alertcolor2}{rgb}{0.8, 0, 0}

\makeatletter
\def\BState{\State\hskip-\ALG@thistlm}
\makeatother

\def\sharedaffiliation{%
\end{tabular}
\centering
\begin{tabular}{p{\linewidth}}\centering}
\begin{document}

\title{Towards a property graph generator for
  benchmarking}

\numberofauthors{6} 
\additionalauthors{0}
\author{ 
\alignauthor Arnau Prat-P\'erez
\alignauthor Joan Guisado-G\'amez 
\alignauthor Xavier Fern\'andez Salas  
\sharedaffiliation
\affaddr{DAMA-UPC, Universitat Polit\`ecnica de Catalunya } \\
\affaddr{\{aprat|joan|xavierf\}@ac.upc.edu} 
\and
\alignauthor Petr Koupy 
\alignauthor Siegfried Depner 
\alignauthor Davide Basilio Bartolini 
\sharedaffiliation
\affaddr{Oracle Labs} \\
\affaddr{\{petr.koupy|siegfried.depner|davide.bartolini\}@oracle.com} 
}

%

\maketitle

\begin{abstract}
  The use of synthetic graph generators is a common practice among
  graph-oriented benchmark designers, as it allows obtaining
  graphs with the required scale and characteristics. However, finding a graph
  generator that accurately fits the needs of a given benchmark is very
  difficult, thus practitioners end up creating ad-hoc ones. Such a task is
  usually time-consuming, and often leads to reinventing the wheel. In this
  paper, we introduce the conceptual design of DataSynth, a framework for
  property graphs generation with customizable schemas and characteristics. The
  goal of DataSynth is to assist benchmark designers in generating graphs
  efficiently and at scale, saving from implementing their own generators.
  Additionally, DataSynth introduces novel features barely explored so far, such
  as modeling the correlation between properties and the structure of the graph.
  This is achieved by a novel property-to-node matching algorithm for which we
  present preliminary promising results.  
\end{abstract}

\section{Introduction}

\begin{figure*}[t]
    \centering
    \includegraphics[width=\linewidth]{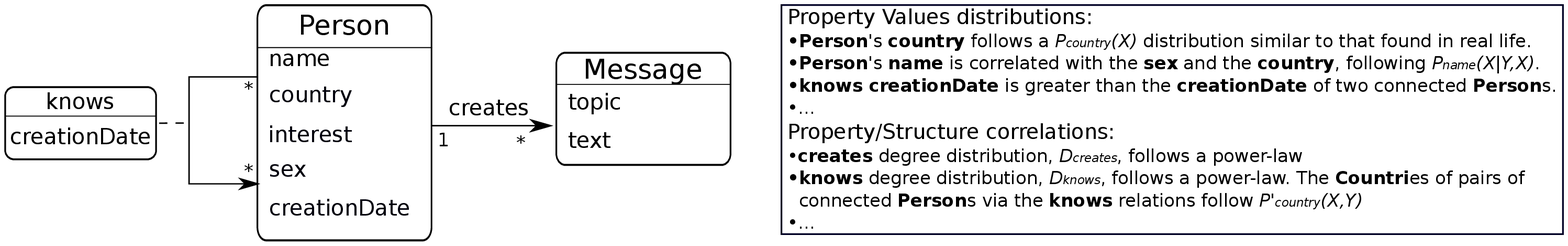}
    \caption{Running Example}
    \label{fig:example}
   \vspace{-0.6cm}
\end{figure*}

During the last decade, the amount of available data has grown exponentially and
it is expected to grow even more over the next years.  Much of these data
present themselves in the form of property graphs, which are graphs whose
vertices and edges are labeled and have associated properties in the form of
key-value pairs.  The increasing popularity of property graphs has provoked the
irruption of many systems specialized on their management~\cite{neo4j,sparksee}
and analysis~\cite{pgx,graphmat,giraph}, as well as benchmarking
initiatives to fairly compare
them~\cite{graph500,ldbcinteractive,ldbcgraphalytics,guo2005lubm,linkbench}.

One of the difficulties of evaluating graph systems is to obtain representative
datasets with the desired scale and characteristics -- because data is often
sensitive and business critical, and companies do not disclose them. Thus, the use of
synthetically generated graphs has become a common practice among graph-oriented
benchmark designers. 

Recent literature on graph system benchmarking reveals an increasing interest on
large synthetic graphs that can reliably mimic real datasets at both the structural and
property value levels. On the one hand, it is acknowledged that the structure of
a graph can heavily affect the performance and behavior of an
algorithm~\cite{prat2011social,wei2016speedup}.  On the other hand, graph-based
technology is penetrating into domains such as social networks, mobility planning
or drug development, just to cite a few.  Each domain requires
application specific benchmarks with graphs where not only the structure is
relevant (i.e. it must be similar to that of the graphs of the domain), but also
the distributions of the property values and the way these properties are
correlated with the underlying graph structure. In many real graphs, we observe
property-structure correlations in the form of joint probability  distributions
between the property values of pairs of connected nodes~\cite{homophily}. The
presence of these property-structure correlations can be determinant for the
performance of some queries. This is an aspect accurately modeled, for instance,
by the modern LDBC Social Network Benchmark~\cite{ldbcinteractive}, which uses
correlated graphs that have been crafted after a detailed choke point analysis
similar to that done on more traditional benchmarking such as
TPC-H~\cite{boncz2013tpc}. 

Given these trends, we foresee an increasing need for synthetic graph
generators that can produce large graphs that are realistic both structurally
and in terms of properties. However, most of existing graph generators only
focus on the structure~\cite{rmat,darwini,bter}, and those that generate
properties are designed for specific use cases~\cite{ldbcinteractive, yutime}.
Implementing a property graph generator is a time consuming task,
thus we need tools to save practitioners from such a burden. 

In this paper we present the conceptual design of DataSynth, a work-in-progress
domain agnostic graph generation framework, for the generation of property
graphs for benchmarking at a scale. DataSynth assumes a shared-nothing
environment and borrows techniques from existing tools to generate data
efficiently in parallel. At the core of DataSynth lies a novel and fundamental
property-to-node matching algorithm that allows decoupling the generation of
properties from the generation of the graph structure, while preserving
property-structure correlations. According to our first experiments, this
approach looks promising.  Summarizing, DataSynth is designed to be capable
of:

\begin{itemize}[leftmargin=*]
		\vspace{-2.4mm}
	\itemsep-0.25em
	\item Generating property graphs using configurable schemas that consist of 
		multiple node and edge types and \mbox{properties}.
	\item Reproducing user-provided property value distributions and 
		property-structure correlations, similar to
	those observed in many real graphs.
\item Scaling to billion edge graphs
	and be work-efficient by applying in-place data generation and other
	optimization techniques whenever possible.
\end{itemize}

The rest of this paper is structured as follows. In
Section~\ref{section:requirements}, we identify the requirements of a
property graph generator and in Section~\ref{section:related} we review the
related work. In Section~\ref{section:concept}, we introduce the conceptual
design of DataSynth. Finally,
in Section~\ref{section:discussion} we conclude the paper.

\section{Graph Generator Requirements}\label{section:requirements}

In this section, we identify and classify data requirements related to the
\textbf{schema}, the \textbf{structure} and the property \textbf{distributions};
and functional requirements related to the \textbf{scale factor} of the graph
and \textbf{other} characteristics of a property graph generator. We
use the running example of Figure~\ref{fig:example}, which represents a simple
social network.

\vspace{-2.5mm}
{\flushleft \textbf{Schema.}}  Graph-based algorithms are being adopted in many
domains, from recommender systems to social
network analysis, route planning, etc. All these
applications rely on property graphs that exhibit a myriad of different schemas
that graph generators have to be able to reproduce. Following the typical
property graph model, such schemas are usually defined in terms of the
\textit{node} and \textit{edge} types, their associated \textit{properties} and
the \textit{cardinality} of the edge types. Thus, a property graph generator
should allow expressing a schema in similar terms.

For instance, in our running example there are two node types, \texttt{Person}
and \texttt{Message}, and two edge types, \texttt{knows}
and \texttt{creates}. \texttt{Person} has five properties: \texttt{name},
\texttt{country}, \texttt{interest}, \texttt{sex} and \texttt{creationDate}.
\texttt{Message} has two: \texttt{topic} and \texttt{text} and the edge
\texttt{creates} has one: \texttt{creationDate}.  Without loss of generality, we
will assume that all properties in this schema are of type String.  Regarding
the cardinality of the edges, \texttt{knows} is a *$\rightarrow$* relationship
between \texttt{Persons}  and \texttt{creates}, which is a 1$\rightarrow$*
relationship between \texttt{Persons} and the \texttt{Messages}.
\vspace{-0.1cm}
{\flushleft \textbf{Structural.}} Graph theory defines tens of structural
properties to characterize graphs, such as number of connected components,
clustering coefficient, degree distribution, centrality, diameter,
assortativity, community distribution, etc. Graphs from different domains
exhibit differences in such structural characteristics, which can affect the
performance of the algorithms. Thus a property graph generator should be able to
generate graphs reproducing them. 

For instance, our running example imposes a structural characteristic, namely
$\mathcal{D}_{knows}$, over the \texttt{knows} edge that should be able to
reproduce.  In addition, the pairs of \texttt{countri}es of connected
\texttt{Person}s by this edge should follow a joint probability distribution
\textit{P$_{country}$(X,Y)} that reflects that \texttt{Person}s from the same
country are more likely to know each other. As a consequence, the resulting
graph will be divided into communities of \texttt{Person}s from the same
\texttt{Country}.
\vspace{-0.1cm}
{\flushleft \textbf{Distribution.}} The distribution of property values in real
graphs is rarely uniform. For example, our running
example, \texttt{Person}'s \texttt{country} should follow a distribution similar
to that found in real life.
Moreover, property values may be correlated with each other. For instance, the
\texttt{name} of a \texttt{Person} is clearly correlated with the \texttt{sex} and
the \texttt{country}.  Finally, other relations may exist between the
values of different properties, such as binary logical relations between
numerical values. For instance, in our running example, the \texttt{knows
  creationDate} should be greater than the \texttt{creationDate} of two connected
\texttt{Person}'s by means of the edge.

\begin{table*}[t]
	\centering   
\scriptsize
 \resizebox{\linewidth}{!}
{
	\begin{tabular}{l|c|c|c|c|c|l|c|c|c|c|c|c|c|c|}
	   \cline{2-15}
\multicolumn{1}{c|}{} & \multicolumn{5}{c|}{Schema} & \multicolumn{1}{c|}{\multirow{3}{*}[\normalbaselineskip]{Structure}} & \multicolumn{2}{c|}{Distributions} & \multicolumn{3}{c|}{Scale Factor} & \multicolumn{3}{c|}{Others} \\ \cline{2-6} \cline{8-15} 
\multicolumn{1}{c|}{} & \multicolumn{2}{c|}{Node} & \multicolumn{2}{c|}{Edge} & Edge & \multicolumn{1}{c|}{} & \multicolumn{1}{c|}{\begin{tabular}[c]{@{}c@{}}Property\\ values\end{tabular}} & \multicolumn{1}{c|}{\begin{tabular}[c]{@{}c@{}}Property\\ Structure\end{tabular}} & \multirow{2}{*}{Node} & \multirow{2}{*}{Edge} & \begin{tabular}[c]{@{}c@{}}Node \\ +\end{tabular} & \multirow{2}{*}{Scalability} & \multirow{2}{*}{Language} & \multirow{2}{*}{Integrability} \\
 & type & prop. & type & prop. & cardinality & \multicolumn{1}{c|}{} & \multicolumn{1}{c|}{distribution} & \multicolumn{1}{c|}{correlation} &  &  & edge &  &  &  \\ \hline
\multicolumn{1}{|l|}{LDBC-SNB~\cite{ldbcinteractive}} &  &  &  &  & x & dd, cc & \multicolumn{1}{c|}{x} & \multicolumn{1}{c|}{} &  &  & x & x &  & x \\ \hline
\multicolumn{1}{|l|}{Myriad~\cite{myriad}} & x & x & x &  & x$^1$ & dd &  &  & x &  &  & x &  &  \\ \hline
\multicolumn{1}{|l|}{RMat~\cite{rmat}} &  &  &  &  &  & pl dd &  &  & x &  &  & x &  &  \\ \hline
\multicolumn{1}{|l|}{LFR~\cite{lfr}} &  &  &  &  &  & pl dd, c &  &  & x &  &  &  &  &  \\ \hline
\multicolumn{1}{|l|}{BTER~\cite{bter}} &  &  &  &  &  & dd, accd &  &  & x &  &  & x &  &  \\ \hline
\multicolumn{1}{|l|}{Darwini~\cite{darwini}} &  &  &  &  &  & dd. ccdd &  &  & x &  &  & x &  &  \\ \hline 
	\end{tabular} 
}
\vspace{-0.2cm} 
\caption{Related work summary. In Structure, dd: degree distribution, cc:cluster coefficient, pl: power law, c: 
communities, accd: average clustering
coefficient per degree, ccdd: clustering coefficient
distribution per degree. x$^1$: supports 1$\rightarrow$1 \& 1$\rightarrow$*. }
\label{table:stateOfTheArt}
\vspace{-0.5cm} 
\end{table*}
\vspace{-0.1cm}
{\flushleft \textbf{Scale Factor.}} Existing benchmarks usually define some sort of 
scale factors for their data. Each scale factor is used to size the 
capabilities of the systems with respect to the amount of processed data. Some 
existing benchmarks base such scale on the number of nodes of the 
graph~\cite{graph500}, others prefer the number of edges~\cite{guo2005lubm} or a 
combination of nodes and edges~\cite{ldbcgraphalytics}, or even on the size on 
disk of the datasets~\cite{ldbcinteractive}. Thus, a property graph generator 
should provide different means of specifying the scale of the produced graph.  
{\flushleft \textbf{Other requirements.}} 
We have identified a series of other characteristics that we believe any graph
generator should have. Specially relevant is its scalability and efficiency,
which have to allow the generator to produce large graphs, as those found in
real-life. Also, taming a cross-domain property graph generator with such a degree of
flexibility requires a properly designed interface. This should include some
sort of Domain Specific Language (DSL) for the specification of the data to
generate, with the corresponding syntax completion tools. Finally, beside the
interface, generators should provide connectors for integrating the framework
with production-level technologies such as databases and cluster storages (e.g.
HDFS).

\vspace{-0.1cm}
\section{Related Work}\label{section:related}

Table~\ref{table:stateOfTheArt} summarizes the state of the art generators in
relation to the requirements described in Section~\ref{section:requirements}.
Note that in the table, marked cells indicates that the generator allows 
configuring explicitly  the corresponding aspect.

The LDBC Social Network Benchmark~\cite{ldbcinteractive} (LDBC-SNB) models a
realistic social network, with multiple node types (Persons, Posts, Topics,
etc.) and edge types (knows, creates, has). Among the novel features it
incorporates, is specially remarkable the generation of a friendship graph with
property-structure correlations, which also has several desirable properties
observed in social networks such as a realistic community
structure~\cite{prat2014community}, a small diameter, a large clustering
coefficient and a Facebook-like degree distribution. However it does not provide
many ways to change the produced schema, but some distributions and
cardinalities can be tuned using configuration files. 

Myriad is a domain-agnostic property graph generator for structured relational
data.  It is flexible and allows the definition of different domain objects with
multiple properties, including foreign keys, which can be seen as one-to-one or
one-to-many edges. However, Myriad does not allow generating many-to-many
relationships, thus cannot be applied to fully model property graphs.
Additionally, Myriad implements in-place data generation using pseudo-random
number generators, a technique we borrow for DataSynth. 

RMat is a graph generator used in the Graph-500
competition~\cite{graph500} and produces graphs with a power-law degree
distributions. Similarly, the LFR
graph generator not only generates power-law degree distributions but also
communities of nodes. This graph generator is typically used to benchmark
community detection algorithms, since the communities are known beforehand. 

The BTER graph generator goes beyond degree distributions and is also capable of
reproducing the average clustering coefficient per degree of an input graph. As
a side effect of its generation process, BTER produces graphs with a positive
degree of assortativity and a community structure.  Darwini~\cite{darwini}
extends BTER and captures the clustering coefficient
distribution at a finer granularity. Both BTER and Darwini are highly scalable,
which allows the generation of Facebook-scale graphs in the order of a trillion
of edges. Additionally, BTER and Darwini produce graphs with a small
diameter due to its generation process, although this cannot be
configured in any way.

\section{DataSynth}\label{section:concept}

\begin{figure}[t]
  \centering
  \includegraphics[width=0.9\linewidth]{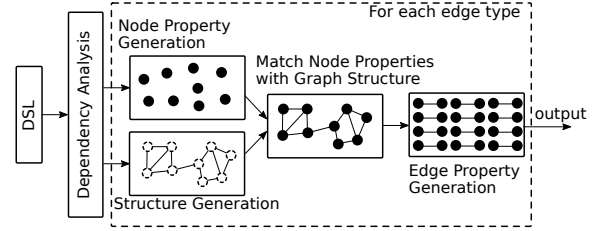}
  \caption{DataSynth general approach}
  \label{figure:architecture}
  \vspace{-0.6cm}
\end{figure}

In this Section, we describe how DataSynth approaches the problem of property
graph generation, given the requirements identified in
Section~\ref{section:requirements}. Figure~\ref{figure:architecture} summarizes
how conceptually DataSynth generates property graphs. First the schema is
received expressed in a domain specific language (DSL), that allows expressing
all the needs identified by the \textbf{schema}, \textbf{structural},
\textbf{distributions} and \textbf{scale factor} requirements. Then, for each
edge type, we generate node
properties and graph structure independently, which are later matched (node ids
are assigned to graph structure nodes) in order to reproduce the required joint
probability distributions specified by the user.  Finally, the properties of the
edges are generated.   

We follow this approach for the following reasons. Building a graph generator
capable of configuring all the existing structural characteristics yet
generating properties at the same time, that also reproduces the joint
probability distributions between property values of nodes, is a very complex
task. Moreover, we do not even know which of the structural characteristics have
an actual impact on the performance of the algorithms, which may actually depend
on the domain of the generated graph and the type of queries to perform. For
instance, while it is acknowledged that the degree distribution and diameter
affect the performance of some algorithms such as BFS, impact of the community
structure or the degree of assortativity is not yet assessed. Thus, our approach
lets the user to choose between existing structure generators and structural
properties they reproduce, fulfilling the \textbf{structural} requirement and
keep the framework open to advances in the field.

In the rest of the section, we detail the whole approach. We first introduce
some preliminary concepts, namely the data model, property generators and 
structure generators, and continue with the actual property graph generation 
process.  We do not detail the design of the DSL because it is not in the scope of 
this paper.

\subsection{Preliminaries}

\vspace{-3mm}
{\flushleft \textbf{Data Model.}} DataSynth is designed with scalability in
mind, for that purpose we rely on distributed tables as the data storage. In
more detail, we use one ``Property Table'' (PT), which is a 2-column table
[id:Long, value:type],  for each pair <node type, property> and <edge type,
property>. In our running example, it would create eight PTs. In addition, we
use one ``Edge Table'' (ET), which is a 3-column table [id:Long, tailId:Long,
headId:Long], for each edge type. The first column is used to identify
an edge instance, while the second and third columns contain the $ids$ of the
nodes connected by the edge.  The $ids$, either of nodes or edges, are unique
per type, and range between 0 and $n-1$, where $n$ is the number of instances of
the given type. For our running example, DataSynth would create three ETs.  

\vspace{-0.1cm}
{\flushleft \textbf{Property Generators (PGs).}} PGs are 
pluggable ``objects'' that can be referenced from the DSL to specify the way
property values are generated. A PG implements an interface with the following methods:

\begin{itemize}[labelsep=-1.5em, leftmargin=*]
  \itemsep-0.25em
  \item \begin{itemize}
		\itemsep-0.25em
		\item [] $initialize: (\dots) \rightarrow void$
		\item [] Sets up the state of the PG. It takes 
			a variable number of parameters that depend on the strategy to 
			generate the data (e.g. a filename to load a dictionary).
	\end{itemize}
\item \begin{itemize}
		\itemsep-0.25em
		\item [] $run : (id : Long, r(id) : Long, \dots) \rightarrow T$
    \item [] It generates the property values of the 
			instances. It takes two parameters, i) the $id$ of the instance, 
      either of a node or an edge, for which it generates the property value and 
      ii) the result of calling $r(id)$ which is a deterministic function that 
      generates a random number using $id$.  Optionally, it also takes a 
			variable number of parameters used to specify the correlation 
      between property values.
	\end{itemize}
\end{itemize}

Notice that $run$ depends exclusively on the $id$ and the result of a
deterministic function called with the same $id$. This allows regenerating a
property value in-place by just knowing the $id$, for instance, in different
computing nodes.  This approach is the same to that used in Myriad, where the
function $r()$ is a pseudo-random number generator (PRNG) with skip seed. Such a
PRNG implements efficiently a method $r : (i : Long) \rightarrow Long$ that
returns the $ith$ random number in a sequence. The PG can use the number
returned by $r()$ to generate a property value randomly. In order to ensure
independence between properties, DataSynth builds a different $r()$
for each PT. Additionally, passing the $id$ to $run$ allows the generation of
user-controlled $uuids$ that can be correlated with other properties such as the
time. 

Note that the interface of a PG is flexible enough to allow the generation of
the properties of our running example. The optional number of parameters of the
\textit{run} method allows implementing the generation of sequences that follow
probability or conditional probability distributions. For example, the
\textit{run} method of $PG_{name}$, the PG for \texttt{Person\ name}, which
depends on \texttt{Person\ country} and \texttt{Person\ sex}, has the signature
$run : (id, r(id), String, String) \rightarrow String$. In order to generate
\texttt{names} with a realistic given distribution, that method can
implement the ``Inverse Transform Sampling'' using the provided random number.
This allows fulfilling part of the \textbf{distribution} requirement.

\vspace{-0.1cm}
{\flushleft \textbf{Structure Generators (SGs).}} Similar to PGs, SGs can be
provided by users to customize the generation of the graph structure (the 
edges), and are referenced from the DSL as well. SGs implement the following interface.

\begin{itemize}[labelsep=-1.5em, leftmargin=*]
  \itemsep-0.25em
  \item 
    \begin{itemize}
        \itemsep-0.25em
        \item [] $initialize: (\dots) \rightarrow void$
        \item [] Initializes the SG, similarly to PG.It
			takes a variable number of parameters that depend on the strategy to 
			build the structure (e.g. a file with an empirical degree distribution). 
    \end{itemize}
  \item 
    \begin{itemize}
        \itemsep-0.25em
        \item [] $run: (n : Long) \rightarrow ET$
        \item [] Generates an ET with the edges of
          a graph of size $n$ (number of nodes). The values of $tailId$ and
          $headId$ range between zero and $n-1$, while the ids of the edges
          range between zero and $m-1$, where $m$ is the ET size,
          which  depends on the generation process. 
      \end{itemize}
  \item 
    \begin{itemize}
        \itemsep-0.25em
        \item [] $getNumNodes: (numEdges : Long) \rightarrow Long$
        \item [] Returns the number of nodes to call the
          method \textit{run(n)} with, such that the resulting ET is of size $n$.  
    \end{itemize}
\end{itemize}

This approach allows accommodating state-of-the-art graph
generators such as BTER or Darwini. Their parameters --in this case the degree
and clustering coefficient distributions-- would be passed to the method
\textit{initialize}. A call to the \textit{run} method would then generate the
 structure for the given number of nodes. Finally, the method
\textit{getNumNodes} would be used to specify the
scale of the graph in terms of the number of edges.

\subsection{Property Graph Generation Process}

The data generation process begins analyzing the schema described by the user to 
reveal dependencies among the data to be generated. In more detail, from the 
dependencies analysis we get a dependency graph, which we traverse to preserve 
the dependencies between the tasks. This guarantees that the required parameter 
are available for each task when we execute it.
There are three different types of tasks: generate property, generate graph and
match graph.  
 
The dependency analysis is required for the following reason. Generate property 
and generate graph tasks take the
size of the task to run as an input (e.g. the number of node instances or the
edges of the graph to generate). These sizes are sometimes given by the size of
the output of another task. For example, imagine that the user of our running
example only defines the scale factor of the graph by means of the number of
\texttt{Person}s to be $N$, but says nothing about the rest of
entities. Thus, how many instances of \texttt{Message} does DataSynth have to
generate? Notice that each \texttt{Message} depends on the number of instances 
of the
edge \texttt{creates} (due to the $1 \rightarrow *$ cardinality). In turn, the
number of edges \texttt{creates} follows $\mathcal{D}_{creates}$, a degree
distribution observed in real-life and provided by the user, and is conditioned
by $N$. Thus, to infer the number of \texttt{Message}s, we need first to
generate the  structure for the edge \texttt{creates}. Once the 
structure of \texttt{creates} is generated, its size determines the number of
\texttt{Message}s to create. Finally, we can apply the match operator between
\texttt{Person}s, \texttt{Message}s and \texttt{creates}. Notice that the chain
of dependencies can be much complex than the one used in this example.

Alternatively, the user could be interested in specifying the scale of the graph
in terms of the number of edges \texttt{creates}, instead of the number of
\texttt{Person}s nor the number of \texttt{Message}s. In this case, DataSynth
would use the \textit{getNumNodes} method with the desired number of edges as a
parameter, and use the result to size the graph structure and the number of
\texttt{Person}s. The size of the resulting graph structure would be used to
determined the number of \texttt{Message}s. This flexibility in specifying the
scale of the graph lets DataSynth to fulfill the \textbf{scale}
requirement.

\vspace{-0.1cm}
{\flushleft \textbf{Generate Structure.}} This task is responsible of the
generation of the graph structure of a given edge type. For each edge type, the
user specifies the SG to use and its parameters. DataSynth initializes the SG
and calls the \textit{run} method, which returns a table with the graph
structure. The number of nodes used to call the run method is determined either
by the user or from the dependency analysis of DataSynth. Remember that this
task generates a graph whose ids must be matched later with node ids to
reproduce the desired correlations (if any).
\vspace{-0.1cm}
{\flushleft \textbf{Generate Property.}} Properties, either from nodes or edges,
are created by calling the \textit{run} method of its corresponding PG $pg$,
which is initialized with the parameters specified by the user using the method
\textit{initialize}. The $run$ method is called $n$ times, which is the size of
the PT $p$ willing to generate.
Before  the generation of $p$, its
corresponding PRNG $r$ is initialized as well. For those properties that are not
correlated with any other property, the $ith$ row of $p$ is
[$i$,$pg.run(i,r(i))$]. For those properties correlated with other
properties is [$i$,$pg.run(i,r(i), val_{0},\dots,val_{k})$], where $val_{j}$
is the result of calling the \textit{run} method for the generation of the $jth$
property the currently generated property depends on (using the appropriate PG
and PRNG). Such call, in turn, 
can lead to successive calls of other run methods. 
The dependency analysis
guarantees that the recursion terminates. For example, in order to generate the
property \texttt{sex} of \texttt{Person}, we would call:
\begin{align*}
  pg_{sex}.run(i,r_{sex}(i), pg_{country}.run(i,r_{country}(i)))
\end{align*}   
Properties for edges can be similarly generated, using the $ids$ of the
endpoints of the edge if needed. Note that thanks
to our approach where property values are generated independently,
these can be generated efficiently in parallel in a distributed system by just
knowing the $ids$ of the nodes to generate. Knowing such $ids$ is easy, because
we just need the number of instances which are unique per type and not
globally. This allows achieving the desired scalability expressed in the
\textbf{others} requirement.
\vspace{-0.1cm}
{\flushleft \textbf{Graph Matching.}} The task of graph matching consists in
matching entries of a PT $p$ with the nodes of the generated graph
structure $g$, in such a way that the desired property-structure correlation is
preserved.  We model the property-structure correlation as a joint probability
distribution $P(X,Y)$. This distribution, which is provided by the user,
expresses the probability of picking a random edge of the graph and observing
property values $X$ and $Y$ in its endpoints. In those cases where an
edge type is not correlated with any property, the matching is done randomly.                            

The input of the graph matching is: the PT $p$ of the property that is correlated
with the structure, a joint probability distribution $P(X,Y)$ and a graph
structure $g$ . The goal is to find a mapping function $f$ that
maps the node ids of the graph structure to ids of the PT, such that the
observed $P'(X,Y)$ after applying the mapping function $f$ is as close as
possible as $P(X,Y)$. This is the way we fulfill the remaining of the
\textbf{distribution} requirement.

We approach the problem using the \textit{Stochastic Block Model (SBM)}. SBM is
a model used for graphs where there are groups of entities with a given property
value or category (one group per property value). In SBM, for each pair of
groups $<i,j>$, there is a probability $\delta_{i,j}$ that an edge exists between
each pair of members of the two groups ($i$ and $j$ can be the same). The SBM is
typically used to study community detection algorithms. 

For example, suppose that in our input PT $p$, there are $n$ different property
values.  Let $Q=\{q_{0}, \dots, q_{n-1}\}$ be the frequencies of each of the
values observed in $p$ (which are the sizes of each group). Let $W$ be a $ n\times n$ matrix such that $W_{ij}$
contains the number of edges between the nodes of group $i$ and $j$\footnote{We
  work with absolute number of edges instead of probabilities for convenience}.
Given $P(X,Y)$ and the number of edges $m$ of the graph structure, we can
compute $W_{ii} = \frac{2m P(i,i)}{q_{i}(q_{i}-1)}$ and
$W_{ij}=\frac{2m P(i,j)}{q_{i}q_{j}}$ if $i!=j$. In other words, 
our problem is equivalent to classifying the nodes of $g$ 
into $n$ groups of sizes
$Q=\{q_{0}, \dots, q_{n-1}\}$, in such a way that the intra and inter-group
edges are as close as possible to those in $W$. Then, the function $f$ is built
by assigning to each node of $g$ an $id$ out of those of $p$ that have the
value corresponding to the partition the node has been assigned. Thus, our
problem can be seen as a graph partitioning problem.

To solve this graph partitioning problem, we
have implemented a variation of the LDG streaming graph partitioning
algorithm~\cite{stanton2012streaming}. In LDG, a node arrives along with its
edges, and is placed to that partition where lay most of its neighbors already
seen (weighted by a factor depending on the remaining capacity of the partition).
In our case, instead of taking the decision based on the node's degree, we place
the node to the partition $t$ that minimizes the Frobenius Norm between $W_T$ 
and $W$:
\begin{align}\label{equation:sbm}
  \argmin_{t}{\parallel W_t - W \parallel^{2}_{F}}, 
\end{align}
where $W_t$ is the $n\times n$ matrix where $W_{t,ij}$ contains the number of
edges connecting nodes with properties $i$ and $j$, given that we put the
current node to partition $t$. As in LDG, the score is balanced by the remaining
capacity $(1-\frac{s_t}{q_t})$, where $s_t$ is the number of nodes placed to
partition $t$ so far. We name this method as SBM-Part. Notice that a small
variation of SBM-Part can also be applied to bi-partite graphs, since the SBM
can model this type of graphs as well. If the bi-partite graph is between two
different node types, the input would contain two PTs instead of one.
\vspace{-0.1cm}
{\flushleft \textbf{Preliminary evaluation of graph matching.}}
We conducted some preliminary experiments to assess the quality of the proposed
graph matching. We generated a set of graphs using the LFR and RMAT graph
generators. We have configured LFR with an average degree of 20, a maximum
degree of 50, a minimum community size of 10 and a maximum community size of 50,
which are the parameters used in~\cite{lancichinetti2009community}.  The mixing
factor is set to 0.1.  The rest of parameters have been left to their default
values. We have generated graphs of sizes 10k, 100k and 1M nodes.  In the case
of the RMAT, we have used the default parameters. We have generated graphs of
scale 18, 20 and 22.

We partitioned each of the graphs $g$ into $k$ groups representing $k$ different
values, using LDG. The size of the $ith$ group  is $n \cdot
\frac{max(geo(0.4,i), 1/k)}{\sum_{j=1}^{k}{max(geo(0.4,j), 1/k)}}$, where $n$
is the number of nodes of $g$ and $geo$ is a geometric distribution with
parameter $0.4$ in this case. We use a geometric distribution to emulate 
real-life graphs, where groups have different sizes. Then, the nodes 
of the $ith$ partition were assigned the property
value $i$.  Then, we computed our joint probability distribution $P(X,Y)$
empirically. Finally, we created a PT $p$ with $ids$ between $0$ and $n-1$,
containing as many rows with property value $i$ as the size of partition $i$.
Then, we run SBM-Part using $p$, $P(X,Y)$ and $g$. We sent
the nodes to SBM-Part randomly. 

\begin{figure}[!t]
  \begin{minipage}{0.32\linewidth}
    \includegraphics[width=1.0\linewidth]{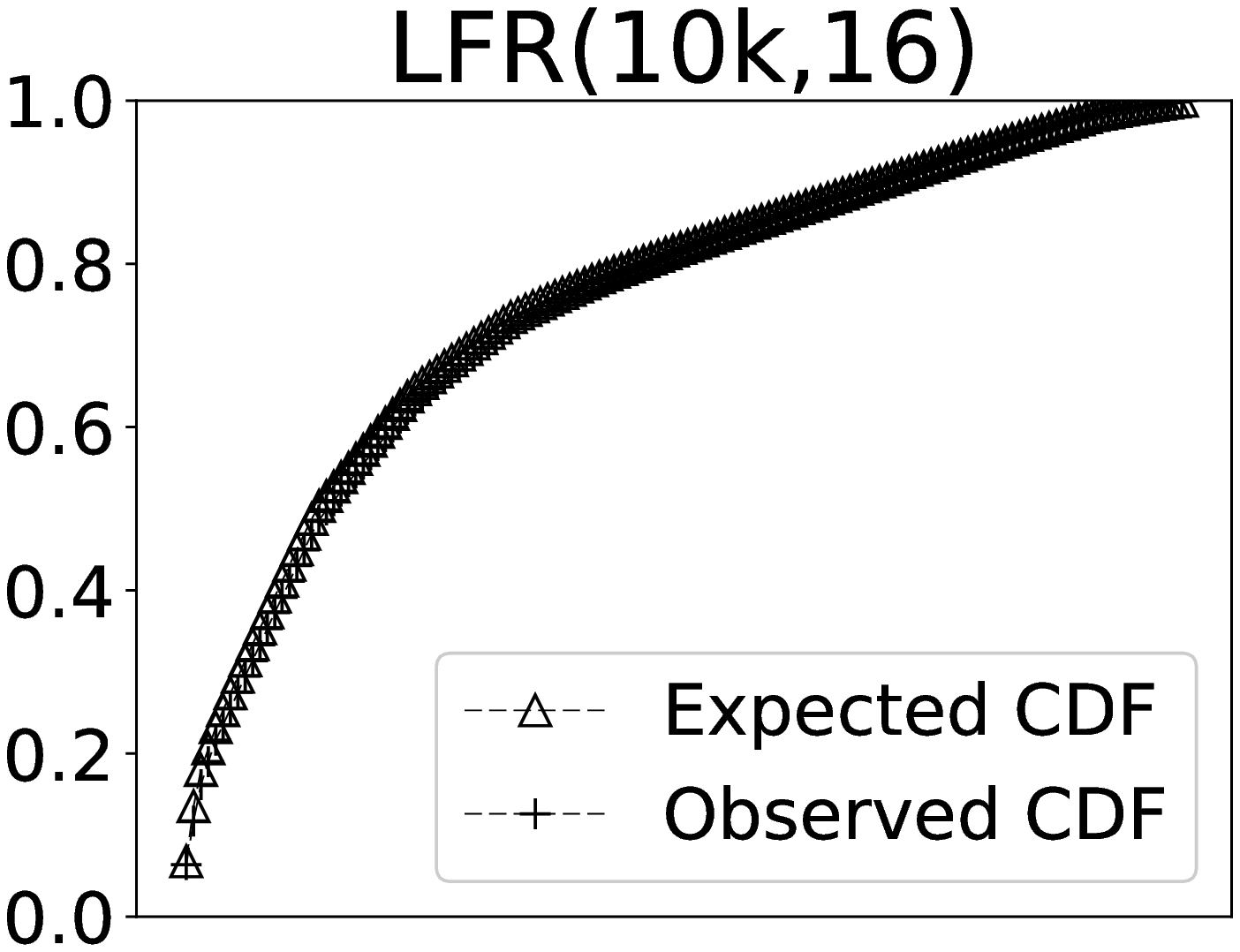}
  \end{minipage}
  \begin{minipage}{0.32\linewidth}
    \includegraphics[width=1.0\linewidth]{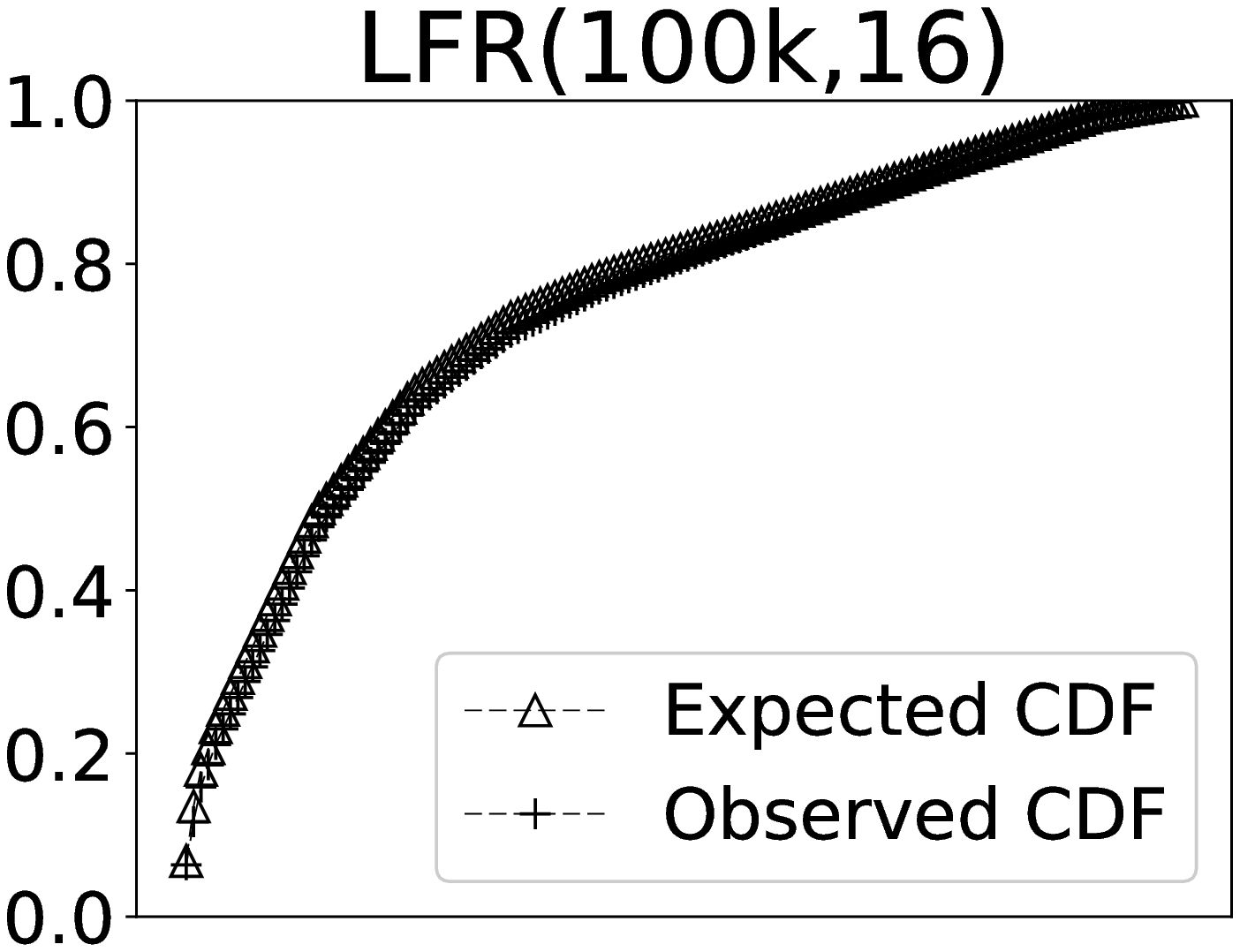}
  \end{minipage}
  \begin{minipage}{0.32\linewidth}
    \includegraphics[width=1.0\linewidth]{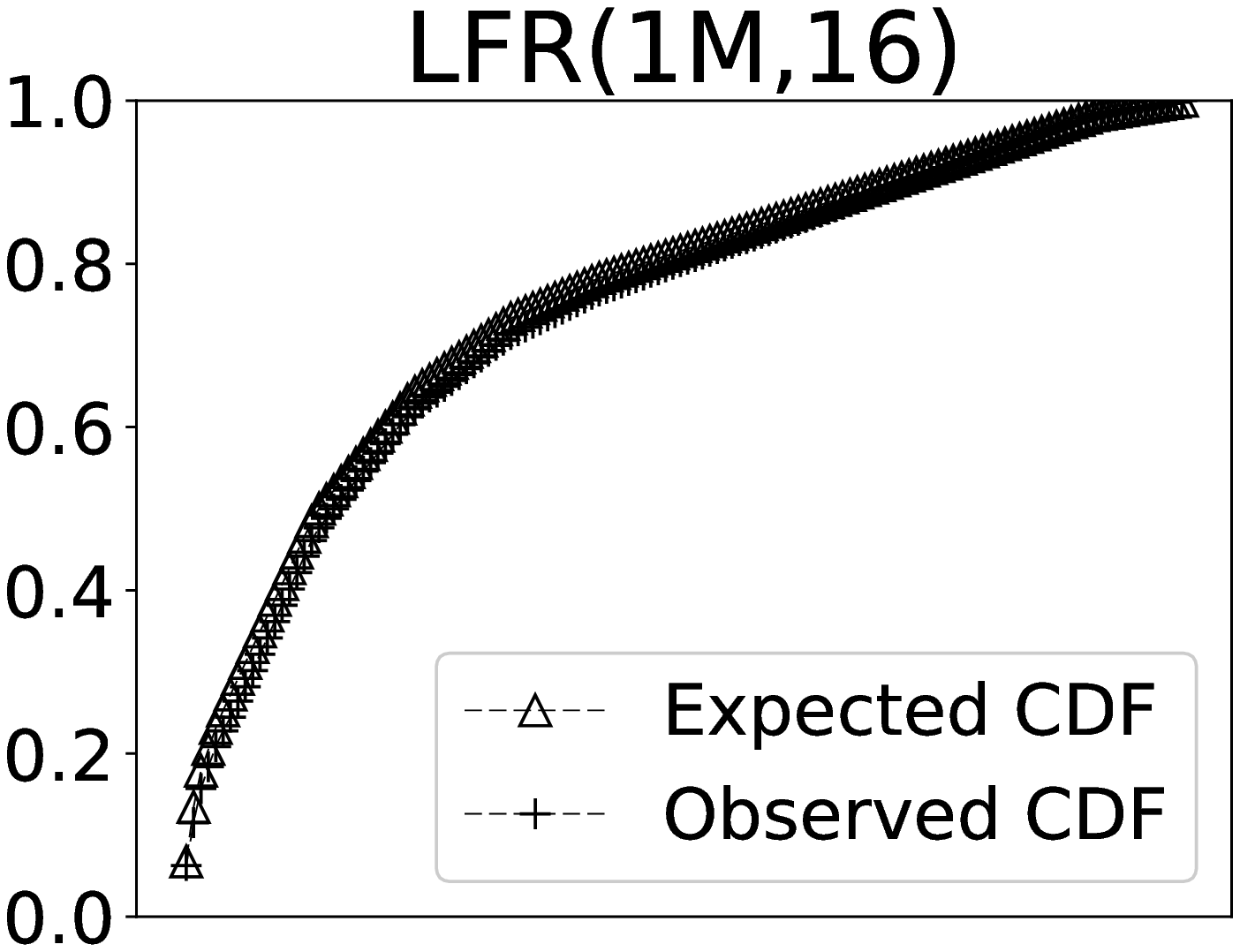}
  \end{minipage}
  \begin{minipage}{0.32\linewidth}
    \includegraphics[width=1.0\linewidth]{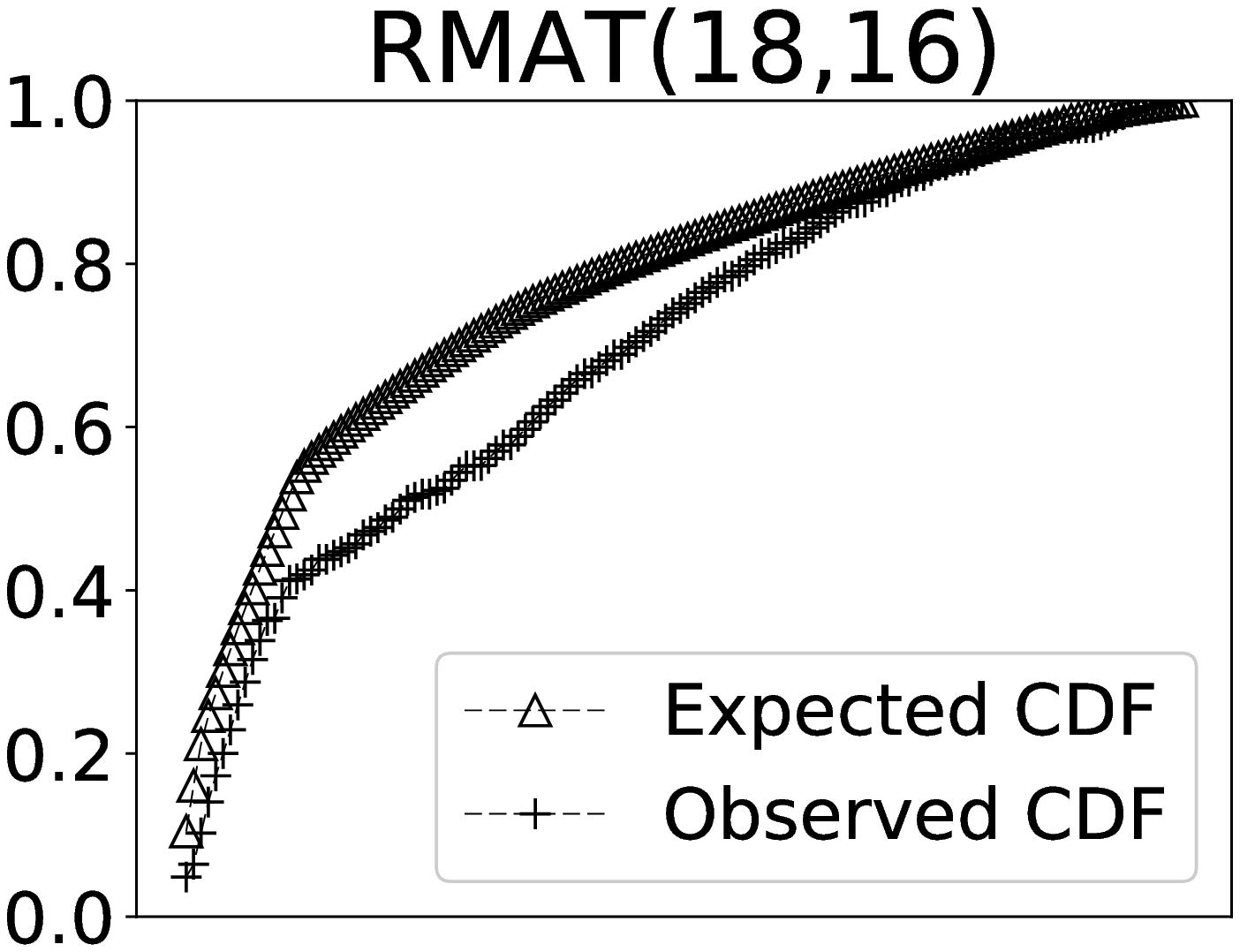}
  \end{minipage}
  \begin{minipage}{0.32\linewidth}
    \includegraphics[width=1.0\linewidth]{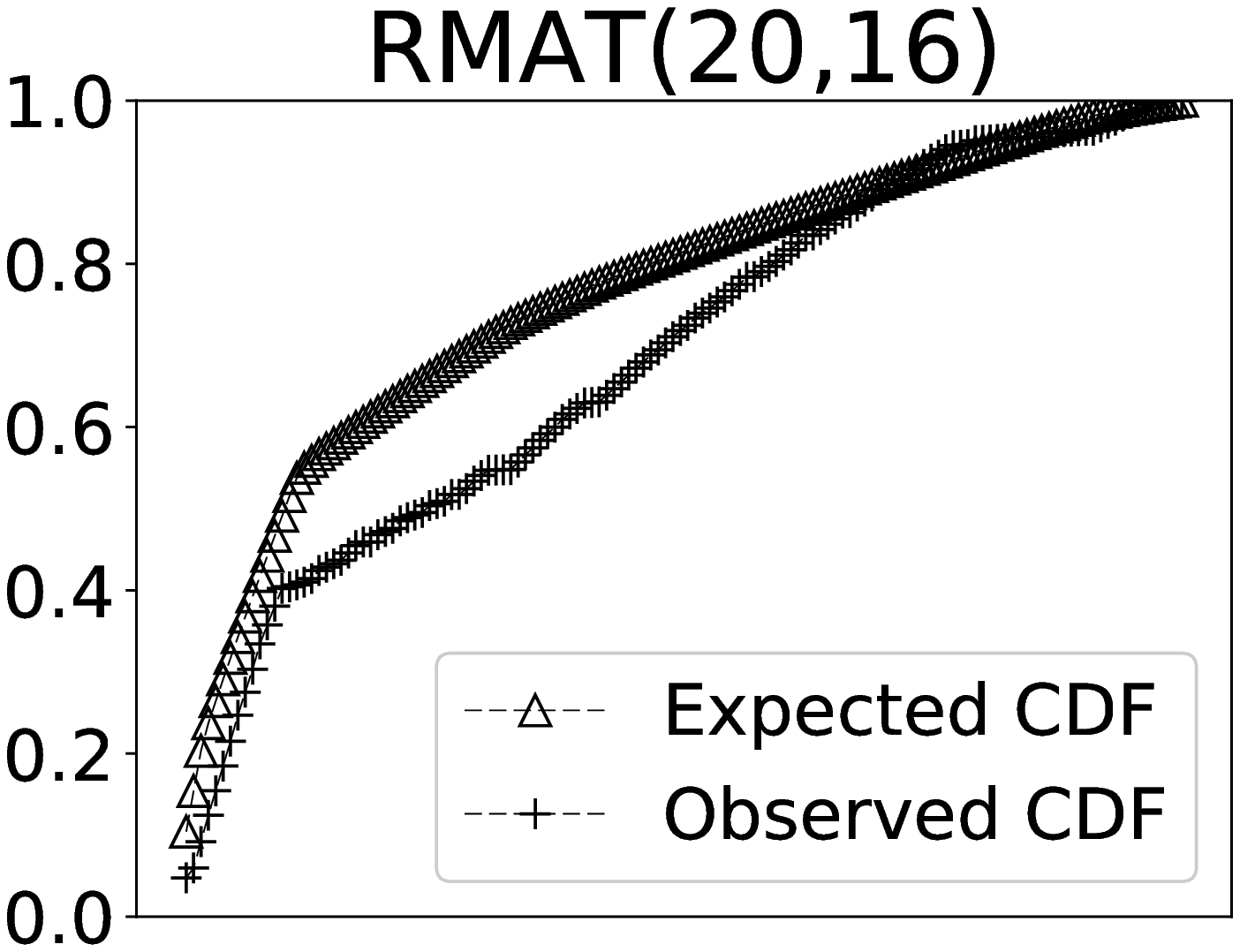}
  \end{minipage}
  \begin{minipage}{0.32\linewidth}
    \includegraphics[width=1.0\linewidth]{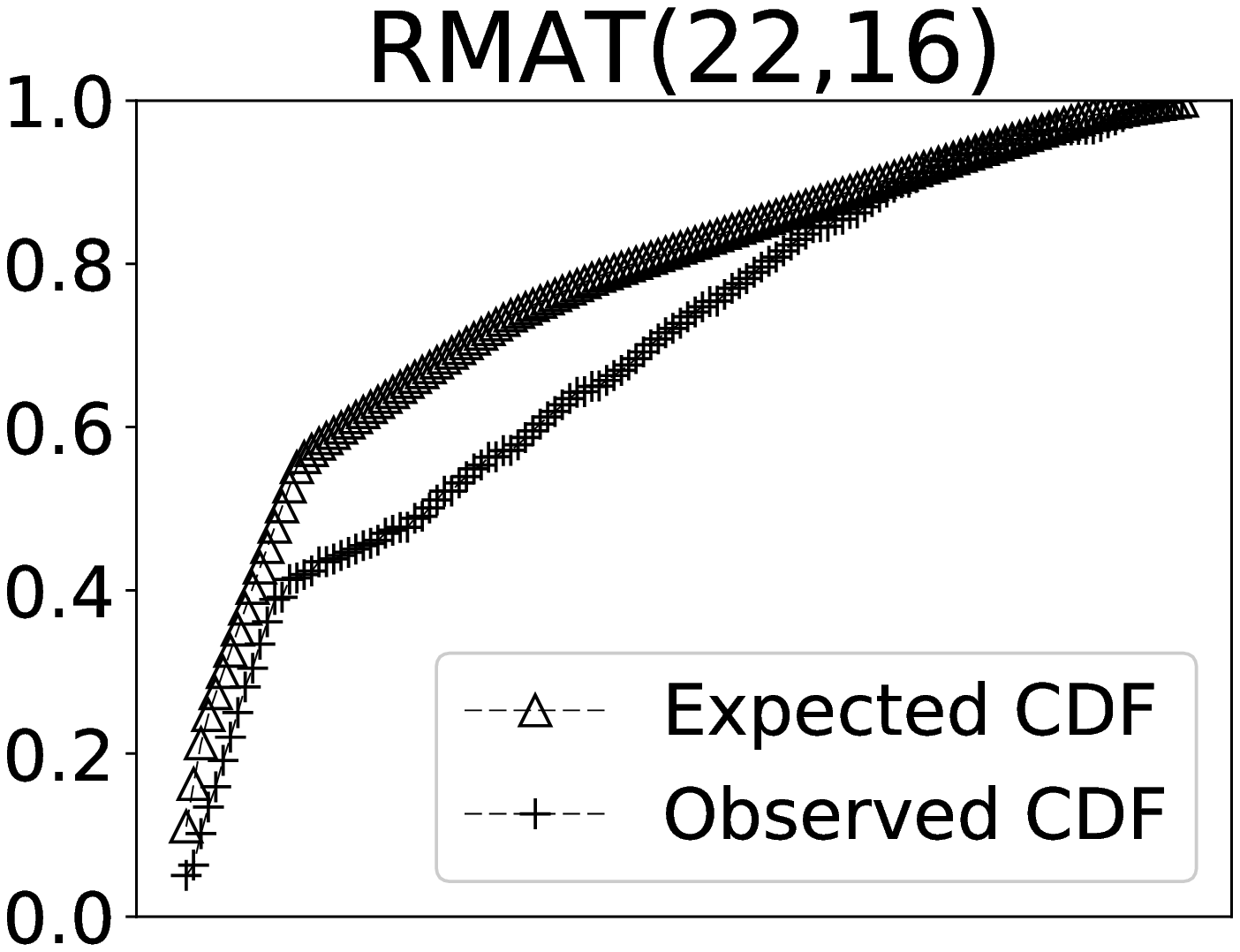}
  \end{minipage}
  \caption{Results for LFR and RMAT graphs of different sizes and 16 different values}
  \label{figure:graphmatchingsize}
  \vspace{-0.3cm}
\end{figure}

\begin{figure}[!t]
  \begin{minipage}{0.32\linewidth}
    \includegraphics[width=1.0\linewidth]{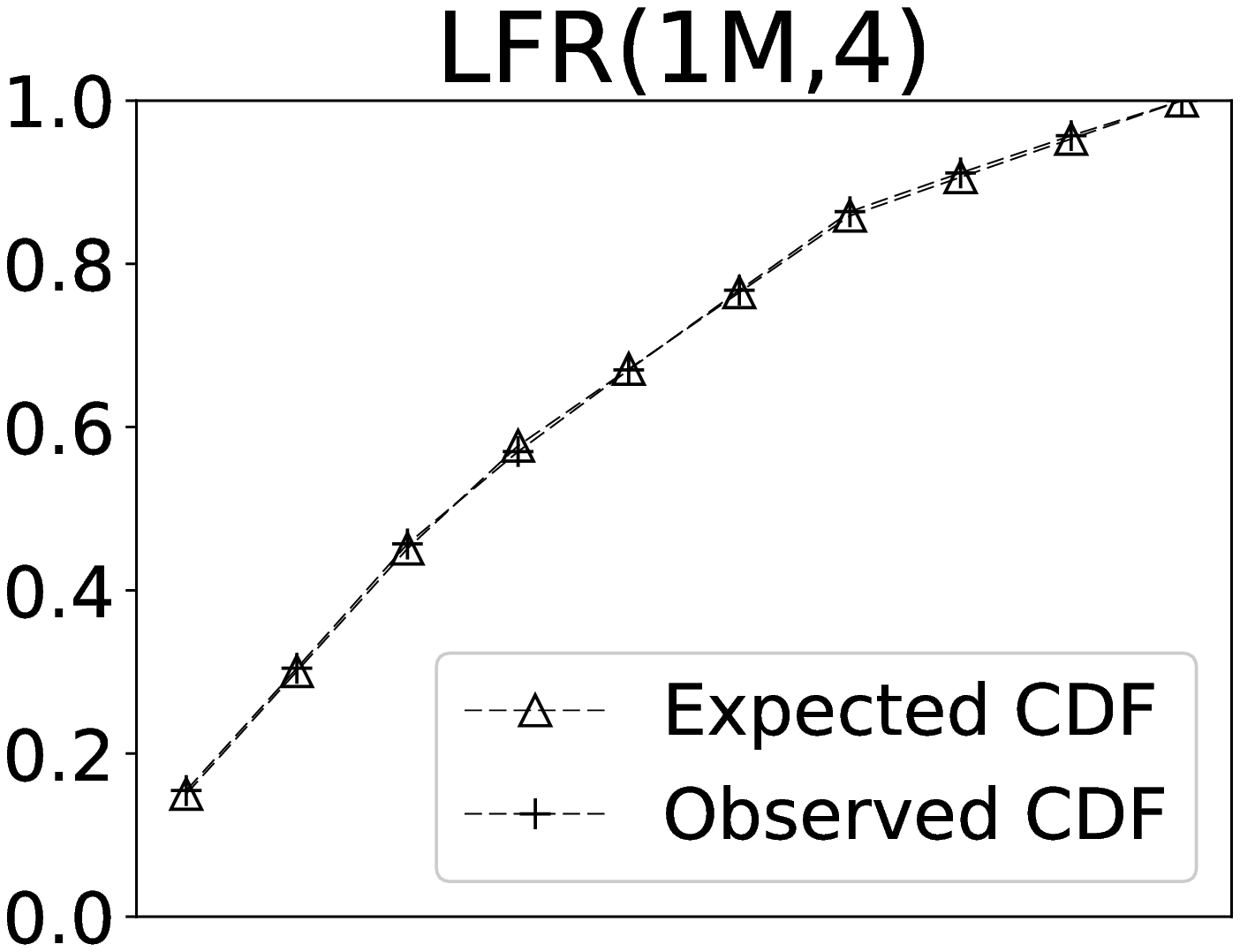}
  \end{minipage}
  \begin{minipage}{0.32\linewidth}
    \includegraphics[width=1.0\linewidth]{./images/lfr/graph_1000000_16.eps}
  \end{minipage}
  \begin{minipage}{0.32\linewidth}
    \includegraphics[width=1.0\linewidth]{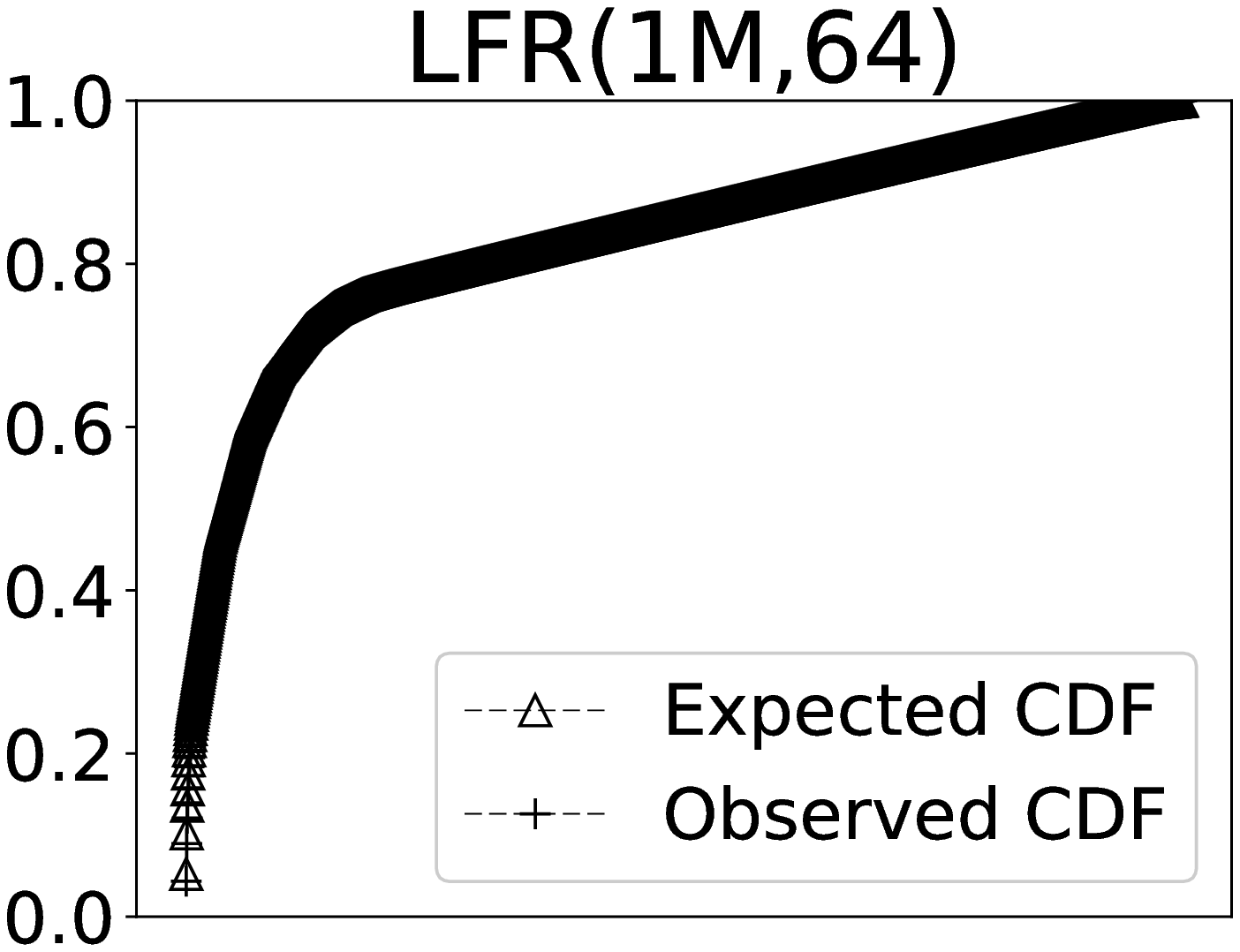}
  \end{minipage}
  \begin{minipage}{0.32\linewidth}
    \includegraphics[width=1.0\linewidth]{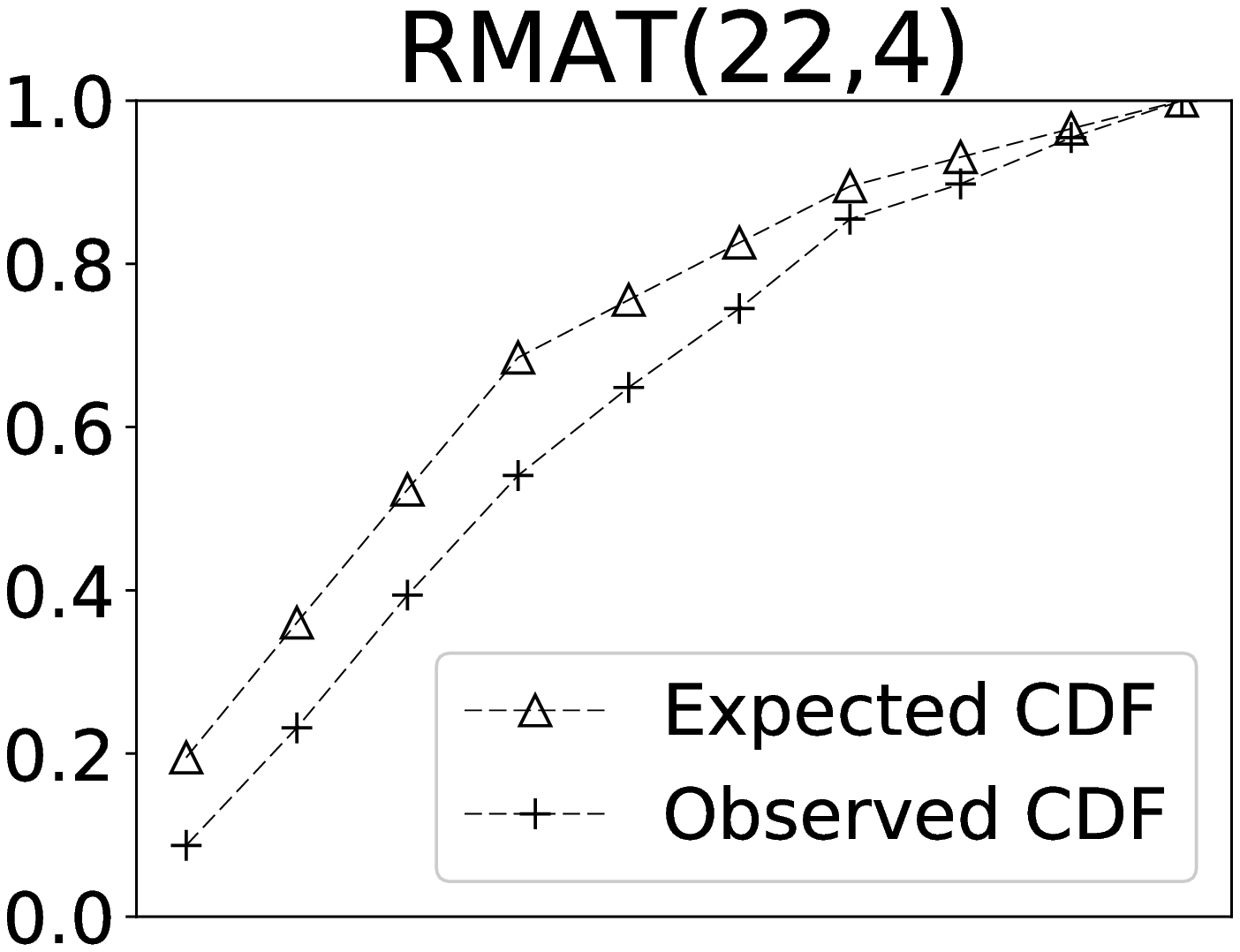}
  \end{minipage}
  \begin{minipage}{0.32\linewidth}
    \includegraphics[width=1.0\linewidth]{./images/rmat/graph_22_16.eps}
  \end{minipage}
  \begin{minipage}{0.32\linewidth}
    \includegraphics[width=1.0\linewidth]{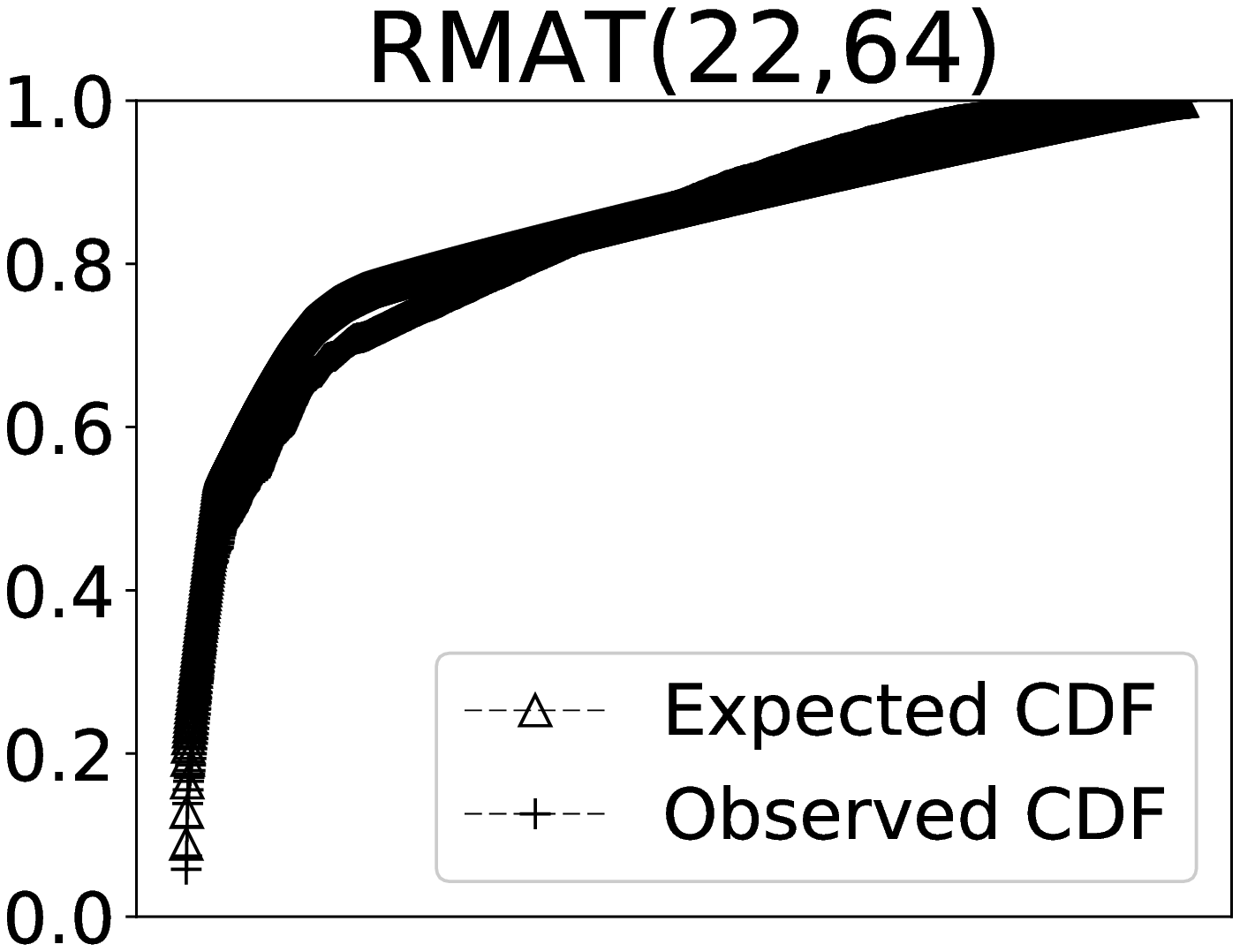}
  \end{minipage}
  \caption{Results for LFR and RMAT graphs of fixed size and different number of values}
  \label{figure:graphmatchingvalues}
  \vspace{-0.5cm}
\end{figure}

Figure~\ref{figure:graphmatchingsize} and~\ref{figure:graphmatchingvalues} show
the CDF of the expected ($P(X,Y)$) and observed $(P'(X,Y))$ distributions after
running SBM-Part for different graphs and number of $k$ values. The x axis
corresponds to the different pairs of values $<i,j>$, and are sorted
by decreasing probability in the expected CDF, for both distributions. 

In Figure~\ref{figure:graphmatchingsize}, we fix the number of $k$ values to 16,
and vary the size of the graph for the two generators. We see two interesting
results. The first is that the quality of the results for LFR graphs seems to be
very good, with the observed distribution with a shape that is very similar to
the expected, and better than that obtained for RMAT graphs. For the latter,
however, note that SBM-Part is able to reproduce the pronounced slope at the
beginning of the distribution, which in general correspond to those entries of
$P(X,Y)$ where $X=Y$. The results suggest that the performance of the algorithm
might be affected by the structure of the graph being partitioned.  The second
result is that the quality of the results does not seem to be affected by the
size of the graphs, which suggests that the method could scale to larger graphs
qualitatively speaking.

In Figure~\ref{figure:graphmatchingvalues}, we see the results when we fix the
sizes and change the number of $k$ to 4, 16 and 64. The results are very similar
to those observed in the previous experiments. The method works consistently
very well with LFR graphs while for RMAT graphs, it seems that the larger the
number of values the better. This seems to confirm that the there is a strong
influence of the structure of the graph to the quality of the results. Finally,
about the performance of the algorithm, it takes about 1100s to process the
largest problem, RMAT-22 (with 67M of edges) and  64 values, using a single
thread on an Intel Xeon E-2630 v3 at 2.4Ghz. No optimizations of any kind have been
implemented.

\section{Discussion and Future Work}\label{section:discussion}

We have presented the design of DataSynth, a
framework for property graph generation with user-defined property distributions
and correlations, different graph structures and property-structure
correlations. The method relies on a novel graph partitioning algorithm, called
SBM-Part, that allows matching properties to nodes in
a graph, in such a way that desired joint probability distributions are
preserved. 

SBM-Part is a greedy algorithm that does not guarantee an optimal solution, thus
strict constraints 
cannot be fully guaranteed.
However, special cases of one-to-one and one-to-many edges could be
efficiently handled by more specific and efficient operators. These, would
generate both the property values and the graph structure at the same time,
which would boost performance allow reproducing strict constraints reliably. 
Other specific graph structures such as trees, which appear in message cascades
in social networks, might require also special strategies. In this case,
information propagates through the cascade, which could be modeled using a
vertex-centric approach that propagates the information through the cascade
iteratively.  

We have presented preliminary results of SBM-Part. Further study is
required, including a complexity analysis,  optimization
strategies, etc. Specially interesting would be
understanding which is the relation between the graph structure and the provided
joint probability distribution (i.e. in which situations the algorithm performs
well and which does not). Additionally, performing experiments for multi-valued
properties would also be interesting.   

Finally, more work is needed regarding scalable graph generators with realistic
structural characteristics. So far, we BTER and Darwini are the structural graph
generators that allow tweaking a larger spectrum of structural characteristics.
Studying which of the characteristics are important for a given domain, and
building scalable graph generators to reproduce these characteristics are still open
problems.


\section*{Acknowledgments}
{
	\setstretch{1.5}

	\scriptsize  DAMA-UPC thanks the Ministry of Economy, Industry 
	and Competitiveness of Spain, Generalitat de Catalunya, for grant numbers 
	TIN2013-47008-R and SGR-1187 respectively and also the EU H2020 for 
	funding the Uniserver  project (ICT-04-2015-688540). Also, thanks to Oracle 
	Labs for the support to our research on graph technologies.
}

\bibliographystyle{plain}
{\scriptsize
\bibliography{main}
}


\end{document}